\documentclass[twocolumns,a4paper]{aa}
\usepackage[comma,authoryear]{natbib}
\usepackage[dvips]{graphicx}
\usepackage[english]{babel}
\usepackage[latin1]{inputenc}
\usepackage{latexsym}
\usepackage{amssymb}

\renewcommand{\textbf}{}
\renewcommand{\bf}{}
\renewcommand{\mathbf}{}

\newcommand{\eg}{\textit{e.g.}}

\newcommand{\third}{$3^\mathrm{rd}$}
\newcommand{\vsini}{v \cdot \sin i}
\newcommand{\kms}{\mathrm{km.s}^{-1}}
\newcommand{\Req}{R_\mathrm{eq}}
\newcommand{\Rp}{R_\mathrm{pol}}

\newcommand{\mod}{\mathrm{mod}}

\renewcommand{\l}{\ell}

\begin{document}

\title{Regular patterns in the acoustic spectrum of rapidly rotating stars}

\author{D. Reese$^{1,2}$ \and F. Lignières$^2$ \and M. Rieutord$^2$}

\institute{
$^1$ Department of Applied Mathematics,
University of Sheffield,
Hicks Building,
Hounsfield Road,
Sheffield S3 7RH, UK \\
$^2$ Laboratoire d'Astrophysique de Toulouse-Tarbes,
Université de Toulouse, CNRS,
14 av. Edouard Belin, 31400 Toulouse, France}

\date{}

\offprints{D. Reese \\ \email{D.Reese@sheffield.ac.uk}}

\abstract
{Rapid rotation modifies the structure of the frequency spectrum of pulsating
stars, thus making mode identification difficult.}
{We look for new forms of organisation for the frequency spectrum that can
provide a basis for mode identification at high rotation rates.}
{Acoustic modes in uniformly rotating polytropic models of stars are computed
using a numerical code that fully takes the effects of rotation
(centrifugal distortion and Coriolis acceleration) into account.  All low-degree
modes, $\l=0$ to $3$, with radial orders $n=1-10$ and $21-25$  for $N=3$
polytropic models and $n=1-10$ for $N=1.5$ polytropic models are followed from
a zero rotation rate up to $59\,\%$ of the break-up velocity.}
{We find an empirical formula that gives a good description of the
high-frequency range of the computed acoustic spectrum for high rotation rates.
Differences between this formula and complete eigenmode calculations are shown
to be substantially smaller than those obtained with a \third\ order
perturbative method valid at low rotation rates.}
{}

\maketitle

\section{Introduction}

Asteroseismology has provided a way of probing stellar interiors based on the
interpretation of observable stellar pulsations.  In order for such
interpretations to be successful, it is necessary to identify pulsation
frequencies by correctly associating them with theoretically calculated
pulsation modes.  This is important because the geometry of a pulsation mode
determines which regions of the star it probes.  One way of doing this is by
using regular patterns which occur in stellar frequency spectra.  While being
quite successful in the case of the Sun and a number of slowly rotating stars
\citep{Michel2006}, mode identification based on pattern recognition has proved
to be very difficult in rapidly rotating stars \citep[\eg][]{Goupil2005}.  One
of the basic reasons for this is that a proper understanding of the structure of
the frequency spectrum has not yet been achieved in the case of such stars.  Up
to now, mainly perturbative methods, valid at low rotation rates, have been
used to evaluate the effects of rotation on pulsation modes and their
frequencies.  While providing a useful context in which to interpret pulsations
of slowly rotating stars, they are unable to correctly predict the structure of
the frequency spectrum in rapidly rotating stars.  This is particularly clear in
Fig.~5 of \citet{Reese2006}, in which perturbative calculations of frequencies
are compared with complete eigenmode calculations for a polytropic stellar model
rotating at $59\,\%$ of the critical angular velocity.  As a result, many stars
remain out of reach for asteroseismology.  This mainly concerns early-type stars
such as $\delta$ Scuti, which can reach projected equatorial velocities
$(\vsini)$ of $200 \,\,\kms$ \citep{Rodriguez2000}, $\beta$ Cephei stars \citep
[$\vsini \lesssim 300 \,\,\kms$,][]{Stankov2005}, and $\zeta$ Oph stars \citep
[$\vsini \lesssim 400 \,\,\kms$,][]{Balona1999}.  Interestingly, the star
$\zeta$ Oph ($\vsini = 380 \,\,\kms)$ has been observed by MOST and
WIRE, thus revealing the presence of 19 different pulsation modes
\citep[][Bruntt, private communications]{Walker2005}.

Recently, \citet{Lignieres2006} and \citet{Reese2006} came up with a numerical
code which overcomes the limitations of perturbative methods and enables one
to accurately evaluate the effects of rapid rotation on stellar pulsations.  By
analysing their results, it is possible to gain a better understanding of the
structure of the frequency spectrum at rapid rotation rates.  In the case of
spherically symmetric stars, the origin of the regular frequency patterns can be
attributed to the integrability of the ray dynamics which asymptotically
describes the acoustic wave propagation \citep{Gough1993}.  However, in the case
of rapidly rotating stars, acoustic ray dynamics is no longer integrable
\citep{Lignieres2006b}, and it is an open question whether or not patterns will
appear in the spectrum of frequencies.  Nonetheless, \citet{Lignieres2006} have
shown that for axisymmetric modes calculated without the Coriolis force, it is
possible to obtain asymptotic patterns in the spectrum of frequencies.

In this paper, using again stellar polytropic models and a complete
non-perturbative computation of their axisymmetric and non-axisymmetric acoustic
modes, we show that regular frequency patterns are present in the acoustic
spectrum and that they can be used to identify modes.

\section{An empirical formula}

In order to obtain frequency patterns for rapidly rotating stars, we start off
with the set of pulsation frequencies computed in \citet{Reese2006} plus
some additional frequencies.  The $n=1-10,\,\,21-25$,
$\l=0-3$, $m=-\l$ to $\l$ modes for $N = 3$ polytropic
models, and the $n=1-10$, $\l=0-3$, $m=-\l$ to $\l$ modes for
$N = 1.5$ polytropic models were followed from a zero rotation rate up to
$59\,\%$ of the break-up velocity.  This enabled us to label the modes in
rapidly rotating stars based on their correspondence with modes in non-rotating
stars.  We looked for patterns in this frequency spectrum and found that the
frequencies approximately obey the following empirical formula in a corotating
frame:
\begin{equation}
\omega_{n,\,\l,\,m} = \Delta_n n + \Delta_{\l} \l + \Delta_m |m| + \alpha^{\pm}
\label{eq:asymptotic}
\end{equation}
In order to obtain frequencies in an inertial frame, one needs to add the
geometrical term $-m\Omega$.

\begin{figure}[thb!]
\includegraphics[width=8.8cm]{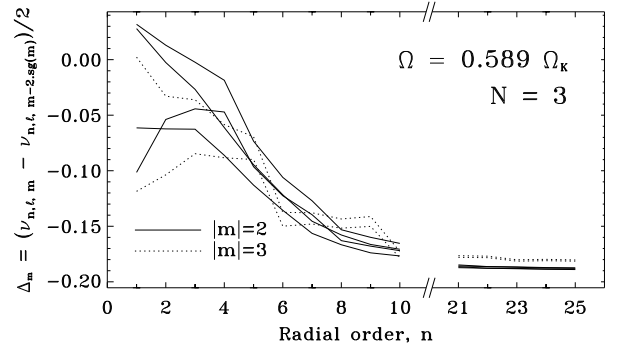}
\caption{Frequency differences $\Delta_m = \frac{\omega_{n,\l,\,m} -
\omega_{n,\l,\,m-2\cdot\mathrm{sg}(m)}}{2}$ as a function of the radial order
$n$ for the $N=3$ polytropic models.  The frequencies have been scaled
by $\Omega_{\star}$.  For high radial orders, these differences converge to a
limit which does not depend on $\l$ or $m$.}
\label{fig:delta_m}
\end{figure}

This formula is a generalisation of Eq.~(42) of \citet{Lignieres2006} by
including the Coriolis effect and the case of non-axisymmetric modes. The terms
$\Delta_n$, $\Delta_\l$, $\Delta_m$ and $\alpha^\pm$, which depend on the
stellar structure, have been computed here for different rotation rates.
The parameter $\alpha^+$ (resp. $\alpha^-$) is an additive constant for symmetric, $\l+m$ even
(resp. antisymmetric, $\l+m$ odd), modes with respect to the equator. We note
that the term for non-axisymmetric modes depends on $|m|$ rather than $m$.
Indeed, at rapid rotation rates or high radial orders, the effects of the
Coriolis force become negligible in comparison with the effects of the
centrifugal force, as can be seen in Fig.~6 of \citet{Reese2006}. The
centrifugal force leads to frequency shifts which do not depend on the sign of
$m$.  In order to know how these shifts scale with $|m|$, we plot in
Fig.~\ref{fig:delta_m} the differences
$\left(\omega_{n,\,\l,\,m}-\omega_{n,\,\l,\,m-2.\mathrm{sg}(m)}\right)/2$ as a
function of the radial order.  We increment the azimuthal order by $2$ rather
than $1$ to insure that the two modes have the same parity. As shown in
Fig.~\ref{fig:delta_m}, these differences converge towards a limit which
slightly depends on $|m|$ as the radial order increases.
This justifies using a linear dependence on $|m|$ as a first
approximation.  The fact that these increments only converge at high radial
orders shows that Eq.~(\ref{eq:asymptotic}) describes an asymptotic behaviour of
the computed acoustic modes.

In Table~\ref{tab:delta}, we give the values of $\Delta_n$, $\Delta_\l$,
$\Delta_m$, $\alpha^+$ and $\alpha^-$, scaled by $\Omega_{\star}$, for
$N=3$ polytropic models.  These are given for different rotation rates
scaled either by $\Omega_{\star}=\sqrt{GM/\Rp^3}$ or by $\Omega_\mathrm{K} =
\sqrt{GM/\Req^3}$, the Keplerian break-up velocity, where $\Req$ (resp. $\Rp$)
is the equatorial (resp. polar) radius. These values were obtained by
calculating various frequency separations and averaging them for radial orders
$21 \leq n \leq 25$.  At zero rotation, they agree pretty well
with the theoretical values, $\Delta_n / \Omega_{\star} = 1.238$ and $\Delta_\l
/\Omega_{\star} = 0.619$ corresponding to the asymptotic description of low
degree and high order acoustic modes in non-rotating stars \citep{Mullan1988}.
The last line contains the same parameters but calculated for frequencies in
which the Coriolis force has been neglected.  The fact that it is essentially
the same as the line before shows that the Coriolis force plays almost no role
in Eq.~(\ref{eq:asymptotic}) nor ultimately in the frequency spectrum for
sufficiently high radial orders.

\begin{table}[htbp]
\begin{center}
\begin{tabular}{*{7}{c}}
\hline
\hline
$\frac{\Omega}{\Omega_\mathrm{K}}$ &
$\frac{\Omega}{\Omega_{\star}}$ &
$\frac{\Delta_n}{\Omega_{\star}}$ &
$\frac{\Delta_\l}{\Omega_{\star}}$ &
$\frac{\Delta_m}{\Omega_{\star}}$ &
$\frac{\alpha^+}{\Omega_{\star}}$ &
$\frac{\alpha^-}{\Omega_{\star}}$ \\
\hline
0.000 &   0.000 &   1.248 &   0.543 &   0.000 &   1.758 &   1.761 \\
0.037 &   0.037 &   1.246 &   0.545 &  -0.007 &   1.757 &   1.764 \\
0.111 &   0.110 &   1.239 &   0.515 &  -0.031 &   1.770 &   1.843 \\
0.186 &   0.181 &   1.234 &   0.395 &  -0.031 &   1.727 &   1.942 \\
0.262 &   0.249 &   1.229 &   0.279 &  -0.036 &   1.680 &   2.006 \\
0.339 &   0.311 &   1.220 &   0.191 &  -0.064 &   1.614 &   2.020 \\
0.419 &   0.368 &   1.194 &   0.189 &  -0.142 &   1.666 &   2.068 \\
0.502 &   0.419 &   1.154 &   0.201 &  -0.182 &   1.661 &   2.036 \\
0.589 &   0.461 &   1.102 &   0.194 &  -0.185 &   1.598 &   1.953 \\
\hline 
0.589 &   0.461 &   1.102 &   0.194 &  -0.185 &   1.587 &   1.943 \\
\hline
\end{tabular}
\caption{Coefficients for Eq.~(\ref{eq:asymptotic}), for $N=3$ polytropes.}
\label{tab:delta}
\end{center}
\end{table}

An important consequence of the values given in Table~\ref{tab:delta}
concerns the small frequency separation.  In non-rotating stars, the so-called
small frequency separation $\omega_{n+1, \ell, m} - \omega_{n, \ell+2, m}$ goes
to zero in the high frequency limit because the ratio $\Delta_n/\Delta_\l$
is $2$.  For rapidly rotating stars, Table~\ref{tab:delta}
clearly shows that this ratio departs from $2$ by taking on larger values.  
This, of course, invalidates the use of the small frequency separation as a mode
identification scheme.

\begin{figure}[htb!]
\includegraphics[width=8.8cm]{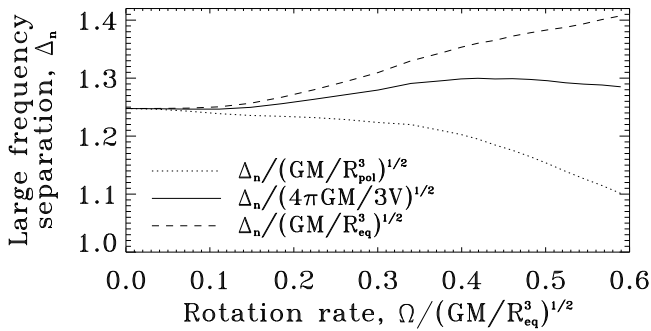}
\caption{The large frequency separation $\Delta_n$ as a function of the rotation
rate $\Omega$.  $\Delta_n$ has been scaled with different quantities so as to
show which other stellar quantity best matches this frequency separation.  As
can be seen, $\Delta_n$ is roughly proportional to the square root of the mean
density of the star.}
\label{fig:delta_n}
\end{figure}

In Fig.~\ref{fig:delta_n}, we plot $\Delta_n$ scaled with different quantities
as a function of $\Omega$.  As can be seen from the figure, $\Delta_n$ is
roughly proportional to $\sqrt{GM/V}$ where $M$ is the mass and $V$ the volume
of the star.  This suggests that $\Delta_n$ may be related in a simple way to
the mean density of the star.

\section{Accuracy of the formula}

\begin{figure}[htb!]
\includegraphics[width=8.8cm]{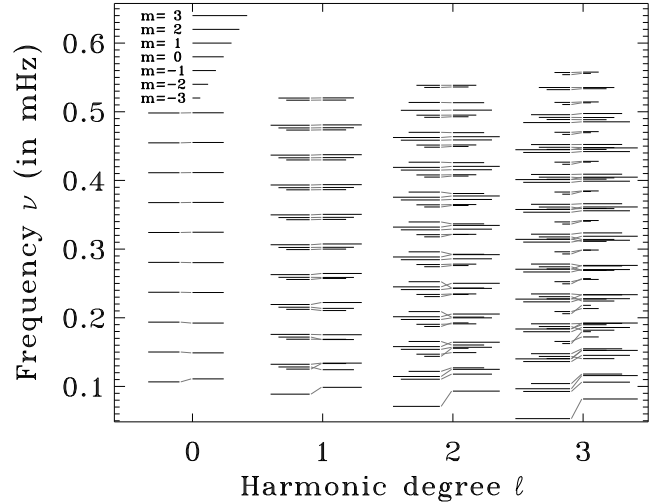}
\caption{Comparison between frequencies based on Eq.~(\ref{eq:asymptotic}) and
complete eigenmode calculations, both of which are given in an inertial frame.
$\l$ and $m$ are indicated on the figure, and the radial orders are $n=1$
to $10$. This figure contains four columns subdivided into two, the left part
corresponding to Eq.~(\ref{eq:asymptotic}) and the right part to complete
calculations.  The oblique grey lines in between show the correspondence
between the two sets of frequencies.  The units are the same as in Fig.~5 of
\citet{Reese2006}, which allows direct comparison.  As can be seen from
comparing these figures, Eq.~(\ref{eq:asymptotic}) gives a much better
description of the frequency spectrum than a \third\ order perturbative
formula.}
\label{fig:comparison}
\end{figure}

We now turn our attention to evaluating the accuracy of the formula
(\ref{eq:asymptotic}).  In Fig.~\ref{fig:comparison}, we plot the correspondence
between a spectrum based on Eq.~(\ref{eq:asymptotic}) and the frequencies given
by the eigenmode calculations for $\Omega = 0.59\,\Omega_K$.  As can be seen,
the two sets of frequencies match pretty well, especially for high radial
orders. This represents a drastic improvement over \third\ order perturbative
calculations (see Fig.~5 of \citealt{Reese2006}).

Table~\ref{tab:error} contains the mean quadratic error $\left< \mathcal{E}
\right>^{1/2}$ both for \third\ order perturbative calculations and empirical
ones.  These errors as well as the probability $p$ of mode inversion (explained
below) have been calculated for modes with ${21 \leq n \leq 25}$ except for the
last line where $1 \leq n \leq 10$.  Physically, an error of $0.0036\,\Delta_n$
at $\Omega = 0.59\,\Omega_K$ corresponds to $0.16\,\,\mathrm{\mu Hz}$ for an
$M=1.9\,M_\odot$, $\Rp=2.3 R_\odot$, $N=3$ polytropic star.  The errors on
the last line can be made smaller by calculating the increments $\Delta_n$,
$\Delta_{\l}$ etc. using radial orders $1 \leq n \leq 10$, which in practise
observers are likely to do.

Another useful quantity to appreciate the reliability of an approximate
formula at mode identification is the probability $p$ of inverting the
identification of two randomly selected frequencies when using such a formula.
This probability is defined as:
\begin{equation}
p = \frac{\mbox{Number of inversions}}{\mbox{Number of pairs of modes}}
\end{equation}
where an inversion occurs for modes $A$ and $B$ when
$\left(\omega_A^\mathrm{approx.} - \omega_B^\mathrm{approx.} \right)
\left(\omega_A-\omega_B\right) < 0$.  $p=1$ corresponds to a spectrum in
reverse order, whereas $p=0$ means the order is identical.  In
Table~\ref{tab:error}, $p$ is calculated for both methods, based on frequencies
in the inertial frame so as to be closer to observations.  For example,
$p=0.0035$ means that out of $3160$ frequency pairs, $11$ were
inverted. As shown, the empirical formula gives a far better idea of the order
in which modes appear in the frequency spectrum at $\Omega = 0.59\, \Omega_K$
than a \third\ order perturbative approach.

\begin{table}[htbp]
\begin{center}
\begin{tabular}{*{5}{c}}
\hline
\hline
& \multicolumn{2}{c}{Perturbative} &
\multicolumn{2}{c}{Asymptotic} \\
$\frac{\Omega}{\Omega_\mathrm{K}}$ &
$\frac{\left< \mathcal{E}^2 \right>^{1/2}}{\Delta_n}$ &
$p$ &
$\frac{\left< \mathcal{E}^2 \right>^{1/2}}{\Delta_n}$ &
$p$ \\
\hline
 0.037 & 0.0005 & 0.0000 & 0.0114 & 0.0000 \\
 0.111 & 0.0370 & 0.0054 & 0.0141 & 0.0044 \\
 0.186 & 0.1793 & 0.0294 & 0.0124 & 0.0022 \\
 0.262 & 0.4325 & 0.0595 & 0.0135 & 0.0047 \\
 0.339 & 0.7436 & 0.0826 & 0.0174 & 0.0041 \\
 0.419 & 0.9924 & 0.1073 & 0.0109 & 0.0013 \\
 0.502 & 1.0674 & 0.1237 & 0.0085 & 0.0025 \\
 0.589 & 0.9885 & 0.1497 & 0.0036 & 0.0035 \\
\hline
 0.589 & 0.3506 & 0.0319 & 0.1235 & 0.0059 \\ 
\hline
\end{tabular}
\caption{Different measurements of the errors for the perturbative and empirical
methods (for $N = 3$).}
\label{tab:error}
\end{center}
\end{table}

As expected, the perturbative method gives best results for low rotation rates,
while Eq.~(\ref{eq:asymptotic}) is more efficient above $\Omega = 0.11\,
\Omega_\mathrm{K}$, for high radial orders. This is because, by
construction, Eq.~(\ref{eq:asymptotic}) preserves the regularity of the
frequency spectrum whereas perturbative methods do not.  Also, comparing the
last two lines of Table~\ref{tab:error} again shows that
Eq.~(\ref{eq:asymptotic}) works better for high frequencies, whereas the
perturbative approach works better for low frequencies, where the centrifugal
force has a smaller effect.

\section{Discussion}

The preceding sections show that at rapid rotation rates, it is possible
to describe the computed frequency spectrum using a formula similar to the
asymptotic one found by \cite{Tassoul1980}.  Quite significantly,
table~\ref{tab:error} shows that the accuracy of this formula
increases at high rotations, thus suggesting that eigenmodes reach an
asymptotic régime at lower radial orders for these rapid rotations.

In the rapid rotation régime, a forthcoming study based on ray dynamics
\citep{Lignieres2008} shows that Eq.~(\ref{eq:asymptotic}) holds for
low-degree, high order modes which concentrate at middle latitudes as rotation
increases. This modification of mode geometry leads to a reorganisation of node
placement as can be seen in Fig.~3 of \citet{Reese2008}, characterised by a
different set of quantum numbers, $(\tilde{n} = 2n + \varepsilon,\tilde{\l} =
\frac{\l - |m| - \varepsilon}{2},\tilde{m} = m)$ where $\varepsilon = (l + m)
\,\,\mod\,\, 2$.  Based on these quantum numbers, Eq.~(\ref{eq:asymptotic}) then
takes on the following form:
\begin{equation}
\omega_{n,\,\l,\,m} = \tilde{\Delta}_{n} \tilde{n} 
+ \tilde{\Delta}_{\l} \tilde{\l} + \tilde{\Delta}_{m} |m| + \tilde{\alpha}^{\pm}
\label{eq:alternate}
\end{equation}
where $\tilde{\Delta}_{n} = \Delta_n/2$, $\tilde{\Delta}_{\l} = 2 \Delta_\l$,
$\tilde{\Delta}_{m} = \Delta_\l + \Delta_m$, $\tilde{\alpha}^+ = \alpha^+$ and
$\tilde{\alpha}^- = \alpha^- + \Delta_\l - \Delta_n/2$.  With the numerical
values from Table~\ref{tab:delta}, we find that $\tilde{\alpha}^+ \simeq
\tilde{\alpha}^-$.  The same also applies for the $N=1.5$ polytropic
model.  This suggests that the true asymptotic formula is closer to
Eq.~(\ref{eq:alternate}) in which $\tilde{\alpha}^+ = \tilde{\alpha}^- =
\tilde{\alpha}$.  Applying this new formula leads to errors which are only
slightly larger in spite of the fact there is now one less free parameter.

Equation~(\ref{eq:alternate}) also accentuates the $|m|$ dependence of the
azimuthal term since $\tilde{\Delta}_m$ is of the same order of magnitude as the
variation on $\Delta_m$.  This suggests that the azimuthal term
$\tilde{\Delta}_m$ expressed in these new quantum numbers should take on a
different form, and calls for further investigation. But as can be seen from
Table~\ref{tab:error}, Eqs.~(\ref{eq:asymptotic}) and~(\ref{eq:alternate}) remain
effective at identifying low-degree (observable) pulsation modes.

For higher degree modes, \citet{Lignieres2006b} pointed out the existence
of other types of modes, namely chaotic and whispering gallery modes, which
correspond, respectively, to intermediate and high values of $\l$ at zero
rotation.  As shown in \citet{Lignieres2008}, the frequency spectra of
these modes have different organisations.

Finally, it must be underlined that the results presented here are based on
polytropic stellar models.  Naturally, the question arises whether or not these
patterns still exist in the case of more realistic models. In the
non-rotating case, sharp chemical gradients lead to the break-down of the
assumptions behind asymptotic analysis.  Nonetheless, rather than removing the
equidistant frequency pattern, these lead to a periodic component which is added
to the asymptotic formula \citep[\textit{e.g.}][] {Gough1990}.  If the rapidly
rotating case behaves similarly, then it may be expected that
these sharp chemical gradients also merely perturb the asymptotic frequency
pattern without removing it altogether.

\section{Conclusion}

The formulae (\ref{eq:asymptotic}) and (\ref{eq:alternate}) that we presented
are the signature of an asymptotic régime of high order acoustic
oscillations in rapidly rotating polytropic stars.  As opposed to perturbative
methods, which are valid at low rotation rates, they give an accurate
description of the {frequency} spectrum by preserving its basic structure
and consequently provide a basis for mode identification schemes. This is a
timely result as seismology space missions are collecting large data sets of
unprecedented quality on rapidly rotating stars.

A key issue for future theoretical studies will be to relate the seismic
observables $\tilde{\Delta}_{n}, \tilde{\Delta}_{\l}$ and $\tilde{\Delta}_{m}$
to the physical properties of the star. Acoustic ray dynamics combined with
semi-classical quantization methods is expected to play a crucial role in this
context.

\begin{acknowledgements}
Many of the numerical calculations were carried out on the Altix 3700 of CALMIP
(``CALcul en MIdi-Pyrénées'') and on Iceberg (University of Sheffield),  both of
which are gratefully acknowledged.
This work was also supported in part by the Programme National de Physique
Stellaire (PNPS of CNRS/INSU) and the Agence Nationale de la Recherche (ANR),
project SIROCO.
\end{acknowledgements}
\scriptsize
\bibliographystyle{aa}
\bibliography{biblio}

\begin{thebibliography}{15}
\expandafter\ifx\csname natexlab\endcsname\relax\def\natexlab#1{#1}\fi

\bibitem[{{Balona} \& {Dziembowski}(1999)}]{Balona1999}
{Balona}, L.~A. \& {Dziembowski}, W.~A. 1999, MNRAS, 309, 221

\bibitem[{{Gough}(1990)}]{Gough1990}
{Gough}, D.~O. 1990, in Lecture Notes in Physics, Berlin Springer Verlag, Vol.
  367, Progress of Seismology of the Sun and Stars, ed. Y.~{Osaki} \&
  H.~{Shibahashi}, 283--+

\bibitem[{{Gough}(1993)}]{Gough1993}
{Gough}, D.~O. 1993, in {Les Houches Summer School Proceedings (1987)}, ed.
  J.-P. {Zahn} \& J.~{Zinn-Justin}, 399--560

\bibitem[{{Goupil} {et~al.}(2005){Goupil}, {Dupret}, {Samadi}, {Boehm},
  {Alecian}, {Suarez}, {Lebreton}, \& {Catala}}]{Goupil2005}
{Goupil}, M.-J., {Dupret}, M.~A., {Samadi}, R., {et~al.} 2005, JApA, 26, 249

\bibitem[{{Ligni{\`e}res} \& {Georgeot}(2008)}]{Lignieres2008}
{Ligni{\`e}res}, F. \& {Georgeot}, B. 2008, submitted

\bibitem[{{Ligni{\`e}res} {et~al.}(2006{\natexlab{a}}){Ligni{\`e}res},
  {Rieutord}, \& {Reese}}]{Lignieres2006}
{Ligni{\`e}res}, F., {Rieutord}, M., \& {Reese}, D. 2006{\natexlab{a}}, \aap,
  455, 607

\bibitem[{{Ligni{\`e}res} {et~al.}(2006{\natexlab{b}}){Ligni{\`e}res}, {Vidal},
  {Georgeot}, \& {Reese}}]{Lignieres2006b}
{Ligni{\`e}res}, F., {Vidal}, S., {Georgeot}, B., \& {Reese}, D.
  2006{\natexlab{b}}, in SF2A-2006: Semaine de l'Astrophysique Francaise, ed.
  D.~{Barret}, F.~{Casoli}, G.~{Lagache}, A.~{Lecavelier}, \& L.~{Pagani},
  479--+

\bibitem[{{Michel}(2006)}]{Michel2006}
{Michel}, E. 2006, Communications in Asteroseismology, 147, 40

\bibitem[{{Mullan} \& {Ulrich}(1988)}]{Mullan1988}
{Mullan}, D.~J. \& {Ulrich}, R.~K. 1988, ApJ, 331, 1013

\bibitem[{{Reese}(2008)}]{Reese2008}
{Reese}, D. 2008, in {HELAS II: Helioseismology, Asteroseismology and MHD
  Connections}, in press

\bibitem[{{Reese} {et~al.}(2006){Reese}, {Ligni{\`e}res}, \&
  {Rieutord}}]{Reese2006}
{Reese}, D., {Ligni{\`e}res}, F., \& {Rieutord}, M. 2006, \aap, 455, 621

\bibitem[{{Rodr{\'{\i}}guez} {et~al.}(2000){Rodr{\'{\i}}guez},
  {L{\'o}pez-Gonz{\'a}lez}, \& {L{\'o}pez de Coca}}]{Rodriguez2000}
{Rodr{\'{\i}}guez}, E., {L{\'o}pez-Gonz{\'a}lez}, M.~J., \& {L{\'o}pez de
  Coca}, P. 2000, A\&A Supp., 144, 469

\bibitem[{{Stankov} \& {Handler}(2005)}]{Stankov2005}
{Stankov}, A. \& {Handler}, G. 2005, ApJS, 158, 193

\bibitem[{{Tassoul}(1980)}]{Tassoul1980}
{Tassoul}, M. 1980, ApJS, 43, 469

\bibitem[{{Walker} {et~al.}(2005){Walker}, {Kuschnig}, {Matthews}, {Reegen},
  {Kallinger}, {Kambe}, {Saio}, {Harmanec}, {Guenther}, {Moffat}, {Rucinski},
  {Sasselov}, {Weiss}, {Bohlender}, {Bo{\v z}i{\'c}}, {Hashimoto},
  {Koubsk{\'y}}, {Mann}, {Ru{\v z}djak}, {{\v S}koda}, {{\v S}lechta}, {Sudar},
  {Wolf}, \& {Yang}}]{Walker2005}
{Walker}, G.~A.~H., {Kuschnig}, R., {Matthews}, J.~M., {et~al.} 2005, ApJ, 623,
  L145

\end{thebibliography}

\end{document}